\documentclass[showpacs,aps,prb,twocolumn,amsmath,amssymb,superscriptaddress]{revtex4}
\usepackage{graphicx}
\usepackage{graphics}
\usepackage{dcolumn}
\usepackage{bm}
\usepackage{amssymb,amsmath}

\begin{document}
\title{Temperature dependence of the paramagnetic spin susceptibility of doped graphene}

\author{A. Faridi}
\affiliation{School of Physics, Institute for Research in
Fundamental Sciences (IPM), Tehran 19395-5531, Iran}
\author{M. Pashangpour}
\affiliation{ Islamic Azad University, Islamshahr Branch, Islamshahr 33135-369, Iran.}
\author{Reza Asgari}
\email{asgari@ipm.ir} \affiliation{School of Physics, Institute for Research in
Fundamental Sciences (IPM), Tehran 19395-5531, Iran}

\begin{abstract}
In this work, we present a semi-analytical expression for the
temperature dependence of a spin-resolved dynamical density-density
response function of massless Dirac fermions within the Random
Phase Approximation. This result is crucial in order to describe
thermodynamic properties of the interacting systems. In
particular, we use it to make quantitative predictions for the
paramagnetic spin susceptibility of doped graphene sheets. We find
that, at low temperatures, the spin susceptibility behaves like
$T^{-2}$ which is completely different from the temperature dependence of the magnetic susceptibility in undoped graphene sheets.
\end{abstract}

 \maketitle

\section{Introduction}

Graphene is a newly realized two-dimensional (2D) electron system
that has attracted a great deal of interest because of the new
physics which it exhibits and because of its potential as a new
material for electronic technology~\cite{reviews,PT}. It exhibits
a large number of new and exotic optical and electronic effects
that have not been observed in other materials~\cite{nair}.

When non-relativistic Coulombic electron-electron interactions are
added to the kinetic Hamiltonian, graphene represents a new type
of many-electron problem, distinct from both an ordinary 2D
electron gas (EG) and from quantum electrodynamics. The Dirac-like
wave equation and the chirality of its eigenstates lead indeed to
both unusual electron-electron interaction effects~\cite{bostwick,
diracgaspapers,barlas_prl_2007,polini_ssc_2007,polini_prb_2008,dassarmadgastheory}
and an unusual response to external
potentials~\cite{tomadin_prb_2008,rossi_prl_2008}.

Spin transport of fermions is central to many fields of physics.
Electron transport runs modern technology and electron spin is
being explored as a new carrier of information~\cite{wolf}. There
has been recent interest in the temperature dependence of various
Fermi liquid properties~\cite{2d-transport}. Technical advances
now allow one to measure the temperature dependence of the
thermodynamic and transport parameters such as  conductivity
at finite temperature in unsuspended~\cite{morozov_prl_2008, chen_nature_nanotech_2008},  and suspended graphene sheets~\cite{bolotin_prl_2008,du_nature_nanotech_2008} and spin susceptibility in two-dimensional systems.

The paramagnetic spin susceptibility shows behavior similar  to
the charge compressibility~\cite{barlas_prl_2007} which decreases as the
interaction increases at zero temperature. This is related to the
fact that the inverse of the spin susceptibility is proportional to
the renormalized Fermi velocity.

Vafek~\cite{vafek_prl_2007} has recently shown that the specific
heat of undoped graphene sheets presents an anomalous
low-temperature behavior displaying a logarithmic suppression with
respect to its noninteracting counterpart as $\sim T/\ln(T)$. Meanwhile, Sheery and
Schmalian studied both the specific heat and the orbital
diamagnetic susceptibility of undoped graphene by using a
renormalization group approach~\cite{sheehy}. They stated that the
dependence of the diamagnetic susceptibility of undoped graphene
on temperature is quite different from the 2D EGs and it behaves like
 $T/|\ln(T)|^2$.

On the other hand, it has been demonstrated
in Refs.~\cite{polini_ssc_2007,polini_prb_2008} (see also Ref.~\cite{dassarmadgastheory}) that doped graphene sheets are normal (pseudochiral) Fermi liquids, with Landau
parameters that possess a behavior quite distinct from
that of conventional 2D EGs. In addition, it was found
that~\cite{ramezan}, at low temperatures, the specific heat has
the usual normal-Fermi-liquid linear-in-temperature behavior, with
a slope that is solely controlled by the renormalized
quasiparticle velocity in doped graphene.

The temperature correction to the spin susceptibility for a 2D EG interacting
via a long-range Coulomb interaction has attracted
a lot of interest over a long period of time~\cite{2d}. It has been shown that the dynamic Kohn anomaly in the density-density response function at $2k_{\rm F}$ and re-scattering of pairs of quasiparticles lead to linear-in-temperature correction to the spin susceptibility~\cite{theory_2d}. Since the static non-interacting density-density response function of doped graphene is a smooth function at $2 k_{\rm F}$ and behaves differently from what one has in standard 2D EG systems, a linear-in-temperature correction to the spin susceptibility does not occur.

In this work we calculate the temperature dependence of a spin-resolved dynamical density-density response function of massless
Dirac fermions within the Random Phase Approximation (RPA) and
subsequently the Helmholtz free energy ${\cal F}(T)$ of doped graphene
sheets where the chemical potential is non zero. This allows us to access important thermodynamic quantities, such as the spin susceptibility which can be
calculated by taking appropriate derivatives of
the free energy. We show  that, at low temperatures, the spin compressibility of doped graphene, in contrary to the diamagnetic spin susceptibility~\cite{sheehy}, behaves as the inverse square of temperature, solely controlled by both the ultraviolet cut-off and graphene's fine-structure constant.

In Sec. II we introduce the formalism that will be used in calculating (paramagnetic) spin susceptibility which includes the many-body effects in the RPA. In Sec. III we present our analytical and numerical results for the free energy and spin susceptibility in doped graphene sheets. Sec. V contains discussions and conclusions.

\section{METHOD AND THEORY }

The agent responsible for many of the interesting electronic
properties of graphene sheets is the non-Bravais honeycomb-lattice
arrangement of carbon atoms, which leads to a gapless
semiconductor with valence and conduction $\pi$-bands. States near
the Fermi energy of a graphene sheet are described by a
spin-independent massless Dirac Hamiltonian~\cite{slonczewski}
\begin{equation}\label{eq:h_kin}
{\cal H} = v_{\rm F} {\bm \sigma}\cdot {\bm p}~,
\end{equation}
where $v_{\rm F}$ is the Fermi velocity, which is
density-independent and roughly three-hundred times smaller that
the velocity of light in vacuum, and ${\bm
\sigma}=(\sigma^x,\sigma^y)$ is a vector constructed with two
Pauli matrices $\{\sigma^i,i=x,y\}$, which operate on pseudospin
(sublattice) degrees of freedom. Note that the eigenstates of
${\cal H}$ have a definite {\it chirality} rather than a
definite pseudospin, {\it i.e.} they have a definite projection of
the honeycomb-sublattice pseudospin onto the momentum ${\bm p}$.

Electrons in graphene do not move around as independent particles.
Rather, their motions are correlated due to pairwise Coulomb
interactions. The interaction potential is sensitive to the
dielectric media surrounding the graphene sheet. The Fourier
transform of the real space potential is given by $v_{q}=2\pi
e^2/\epsilon q $ where $\epsilon$ is the average dielectric
constant between the medium and a dielectric constant of
the substrate.

Within this low energy description, the properties of doped
graphene sheets depend on the dimensionless coupling constant or
graphene's fine-structure constant
\begin{equation}
\alpha_{\rm ee}=  \frac{e^2}{\epsilon \hbar v_{\rm F}}~.
\end{equation}
As it is clearly seen from Eq. (\ref{eq:h_kin}), the spectrum is
unbounded from below and it implies that the Hamiltonian has to be
accompanied by an ultraviolet cut-off which is defined $k_{\rm
c}$ and it should be assigned a value corresponding
to the wavevector range over which the continuum model
Eq.~(\ref{eq:h_kin}) describes graphene. For later purposes we define $\alpha_{\rm gr}= 2g_{\rm v}
\alpha_{\rm ee}$ where $ g_{\rm v}=2$ accounts for valley
degeneracy, $k^{\sigma}_{\rm F}=k_{\rm F}(1+\zeta \sigma)^{1/2}$
is the spin dependent Fermi momentum and $k_{\rm F}=(\pi
n)^{1/2}$ is the Fermi wave number and $\varepsilon_{\rm F}=\hbar v_{\rm F}k_{\rm F}$ is being the Fermi energy with
$n=n^{\uparrow}+n^{\downarrow}$ the total electron density,
$\zeta=(n^{\uparrow}-n^{\downarrow})/n$ is the spin polarization parameter (0 $\leq \zeta \leq$ 1 if we assume that, {\it e. g.}, electrons with real spin $s =\uparrow$ to be majority). For definiteness we take
$k_{\rm c}$ to be such that $\pi k^2_{\rm c}=2(2\pi)^2/{\cal
A}_0$, where ${\cal A}_0=3\sqrt{3} a^2_0/2$ is the area of the
unit cell in the honeycomb lattice, with $a_0 \simeq 1.42$~\AA~the
Carbon-Carbon distance. With this choice, the energy $\hbar v k_{c}=7$ eV and
\begin{equation}
\Lambda =\frac{k_{\rm
c}}{k_{\rm F}}=\sqrt{\frac{2 g_{\rm v}}{n{\cal A}_0}}~.
\end{equation}
The continuum model is useful when $k_{\rm c} \gg k_{\rm F}$, {\it
i.e.} when $\Lambda \gg 1$. Note that, for instance, electron densities $n=0.36 \times 10^{12}$ and $0.36 \times 10^{14}$ cm$^{-2}$ correspond to $\Lambda=100$ and $10$, respectively.

The free energy ${\cal F}={\cal F}_0+{\cal F}_{\rm int}$, a thermodynamic potential at a constant temperature and volume, is usually decomposed into the sum of a noninteracting term ${\cal F}_0$ and an interaction contribution ${\cal F}_{\rm int}$.
To evaluate the interaction contribution to the Helmholtz free energy
we follow a familiar strategy~\cite{Pines_and_Nozieres, Giuliani_and_Vignale} by combining
a coupling constant integration expression for ${\cal F}_{\rm int}$ valid for uniform continuum models ($\hbar=1$ from now on),
\begin{equation}\label{eq:int_free_energy}
{\cal F}_{\rm int}(T)=
\frac{N}{2}\int_{0}^{1}d\lambda\int \frac{d^2{\bm q}}{(2 \pi)^2}v_q\left[
S^{(\lambda)}(q, T) - 1 \right]~,
\end{equation}
with a fluctuation-dissipation-theorem (FDT) expression~\cite{Giuliani_and_Vignale} for the static structure factor,
\begin{equation}\label{eq:structurefactor}
S^{(\lambda)}(q, T) = -\frac{1}{\pi n}\int_0^{+\infty}
d\omega~\coth{(\beta\omega/2)}\Im m \chi^{(\lambda)}_{\rho\rho}(q,\omega, T)~.
\end{equation}
Here $\beta=(k_{\rm B} T)^{-1}$. Quite generally, two-particle correlation functions can be
written in terms of single-particle Green's functions and vertex
parts. The RPA approximation for ${\cal
F}_{\rm int}$ then follows from the RPA approximation for
$\chi^{(\lambda)}_{\rho\rho}(q,\omega)$:
\begin{equation}\label{eq:chi_RPA}
\chi^{(\lambda)}_{\rho\rho}(q,\omega, T) =
\frac{\chi^{\uparrow}_{0}(q,\omega, T)+\chi^{\downarrow}_{0}(q,\omega, T)}{1-\lambda
v_q(\chi^{\uparrow}_{0}(q,\omega,
T)+\chi^{\downarrow}_{0}(q,\omega, T))}
\end{equation}
where $\chi^{\sigma }_{0}(q,\omega, T)$ is the noninteracting
spin resolved density-density response-function in $\sigma$ channel. A central quantity in the many-body techniques is the
noninteracting spin resolved polarizability function $\chi^{\sigma }_{0}(q,\omega, T)$ . The problem of linear density response
is set up by considering a fluid described by the Hamiltonian, which is subject to an external potential. The external potential must be sufficiently weak for low-order perturbation theory to suffice. The induced density change has a linear relation to the external potential through the noninteracting dynamical polarizability function. This function in $\sigma$ channel reads as
\begin{eqnarray}\label{eq:chi_0}
\chi^{\sigma }_{0}(q,\omega, T) &=& g_{\rm v}\lim_{\eta\to
0^+}\sum_{s,s'=\pm}\int \frac{d^2{\bm k}}{(2\pi)^2}
\frac{1 + ss'\cos(\theta_{{\bm k}, {\bm k}+{\bm q}})}{2}\nonumber\\
&\times& \frac{n^{\sigma}_{\rm F}(\varepsilon_{{\bm k}, s}) -
n^{\sigma}_{\rm F}(\varepsilon_{{\bm k}+{\bm q}, s'})}{\omega +
\varepsilon_{{\bm k}, s} - \varepsilon_{{\bm k}+{\bm q}, s'} +
i\eta }~.
\end{eqnarray}
Here $\varepsilon_{{\bm k}, s}=sv_{\rm F}k$ are the Dirac band
energies and $n^{\sigma}_{\rm
F}(\varepsilon)=\{\exp[\beta(\varepsilon - \mu^{\sigma}_0)] +
1\}^{-1}$ is the usual Fermi-Dirac distribution function,
$\mu^{\sigma}_0=\mu^{\sigma}_0(T)$ being the noninteracting
chemical potential. As usual, this is determined by the
normalization condition
\begin{equation}
n^{\sigma} =
\int_{-\infty}^{+\infty}d\varepsilon~\nu(\varepsilon)n^{\sigma}_{\rm
F}(\varepsilon)~,
\end{equation}
where $\nu(\varepsilon) = g_{\rm v} \varepsilon/(2\pi v^2_{\rm
F})$ is the noninteracting density of states. For $T \to 0$ one
finds $\mu^{\sigma}_0(T)=\varepsilon^{\sigma}_{\rm F}\{1-\pi^2
(T/T_{\rm F})^2/6(1+\sigma \zeta)\}$, where $T_{\rm F}=\varepsilon_{\rm F}/k_B$ is
the Fermi temperature. The factor in the first line of
Eq.~(\ref{eq:chi_0}), which depends on the angle $\theta_{{\bm
k},{\bm k}+{\bm q}}$ between ${\bm k}$ and ${\bm k}+{\bm q}$,
describes the dependence of Coulomb scattering on the relative
chirality $s s'$ of the interacting electrons.

After some straightforward algebraic manipulations~\cite{ramezan} we arrive at
the following expressions for the imaginary, $\Im
m~\chi^{\sigma}_{0}$, and the real, $\Re e~\chi^{\sigma}_{0}$, parts
of the noninteracting density-density response function for
$\omega>0$:
\begin{eqnarray}\label{eq:Im_chi_0}
&&\Im m~\chi^{\sigma}_{0}(q,\omega,T)=
\frac{g_{\rm v}}{4\pi}\sum_{\alpha = \pm}\Bigg\{\Theta(v_{\rm F} q-\omega)q^2 f(v_{\rm F}q,\omega)\nonumber\\
&\times&\left[G^{(\alpha,\sigma)}_+(q,\omega,T)-G^{(\alpha,\sigma)}_-(q,\omega,T)\right]\nonumber\\
&+&\Theta(\omega-v_{\rm F}q)q^2f(\omega,v_{\rm F}q)
\left[-\frac{\pi}{2}\delta_{\alpha,-}+H^{(\alpha,\sigma)}_+(q,\omega,T)\right]\Bigg\}\nonumber\\
\end{eqnarray}
and
\begin{eqnarray}\label{eq:Re_chi_0}
&&\Re e~\chi^{\sigma}_{0}(q,\omega,T)= \frac{g_{\rm
v}}{4\pi}\sum_{\alpha = \pm} \Bigg\{\frac{-2k_{\rm
B}T\ln[1+e^{\alpha\mu_0/(k_{\rm B}T)}]}{v^2_{\rm F}}\nonumber\\
&+&\Theta(\omega-v_{\rm F} q)
 q^2f(\omega,v_{\rm F}q)\left[G^{(\alpha,\sigma)}_-(q,\omega,T)-G^{(\alpha,\sigma)}_+(q,\omega,T)\right]\nonumber\\
&+&\Theta(v_{\rm F} q-\omega)q^2f(v_{\rm F}q,\omega)
\left[-\frac{\pi}{2}\delta_{\alpha,-}+H^{(\alpha,\sigma)}_-(q,\omega,T)\right]\Bigg\}~.\nonumber\\
\end{eqnarray}
Here
\begin{equation}
f(x,y)= \frac{1}{2\sqrt{x^2-y^2}}~,
\end{equation}
\begin{equation}
G^{(\alpha,\sigma)}_\pm(q,\omega,T)=\int_{1}^{\infty}du~\frac{\sqrt{u^2-1}}{\exp\left({\displaystyle
\frac{|v_{\rm F} q u\pm\omega|-2\alpha\mu^{\sigma}_0}{2k_{\rm
B}T}}\right)+1}~,
\end{equation}
and
\begin{equation}
H^{(\alpha,\sigma)}_\pm(q,\omega,T)=\int_{-1}^{1}du~
\frac{\sqrt{1-u^2}}{\exp\left({\displaystyle \frac{|v_{\rm F} q
u\pm\omega|-2\alpha\mu^{\sigma}_0}{2k_{\rm B}T}}\right)+1}~.
\end{equation}

\begin{figure}\label{fig:one}
\begin{center}
\includegraphics[height=5.0cm]{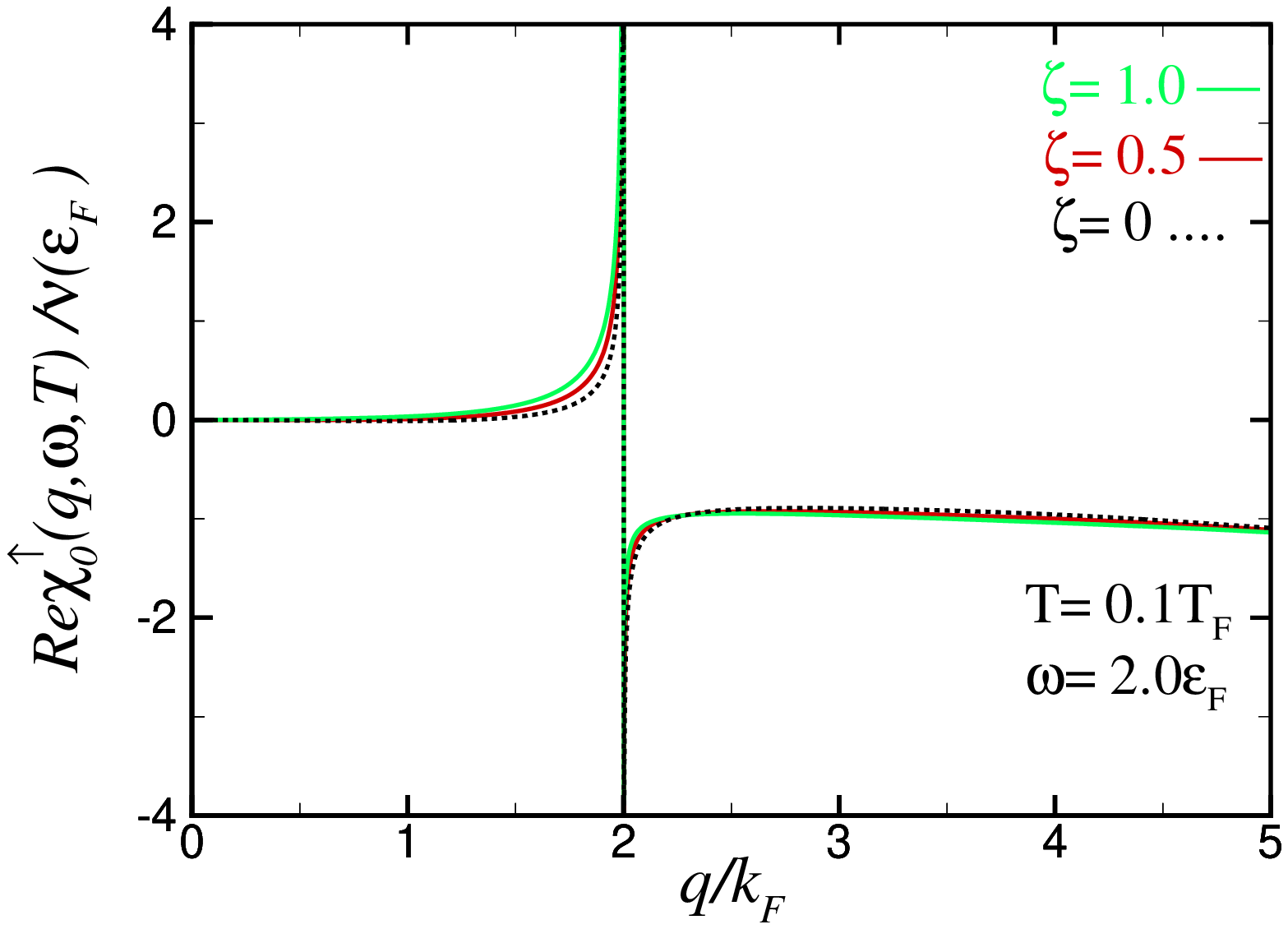}
\includegraphics[height=5.0cm]{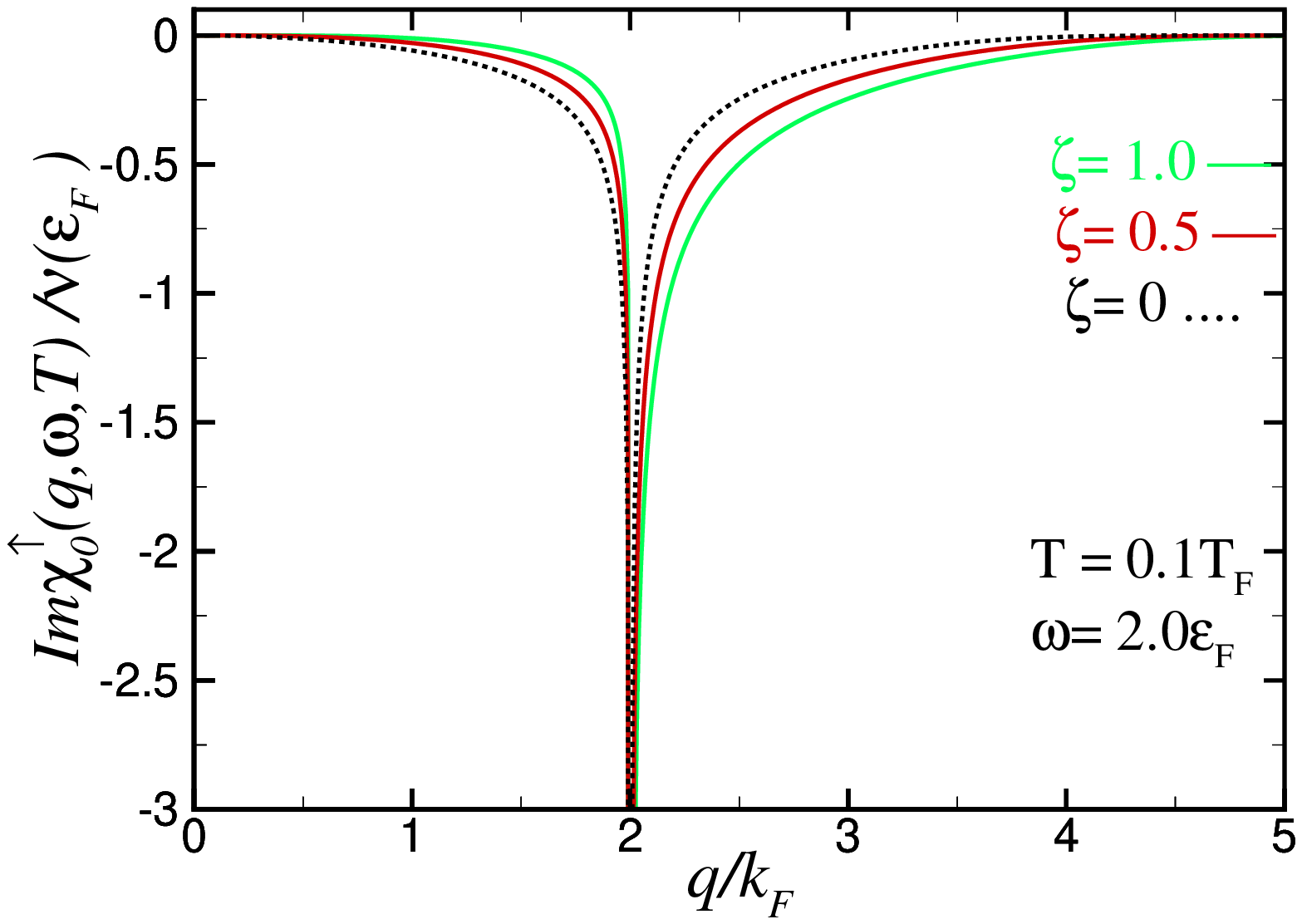}
\caption{( Color online) Upper : the real part of the dynamical response function $\Re e~\chi^{\uparrow}_{0}(q,\omega,T)$
[in units of $\nu(\varepsilon_{\rm F})$] as a function of
$q/k_{\rm F}$ for $\omega=2\varepsilon_{\rm F}$, $T=0.1 T_{\rm F}$ and three values of $0 \leq \zeta \leq 1$.
Bottom: same as in the upper panel but for the imaginary part.}
\end{center}
\end{figure}

\begin{figure}\label{fig:two}
\begin{center}
\includegraphics[width=0.90\linewidth]{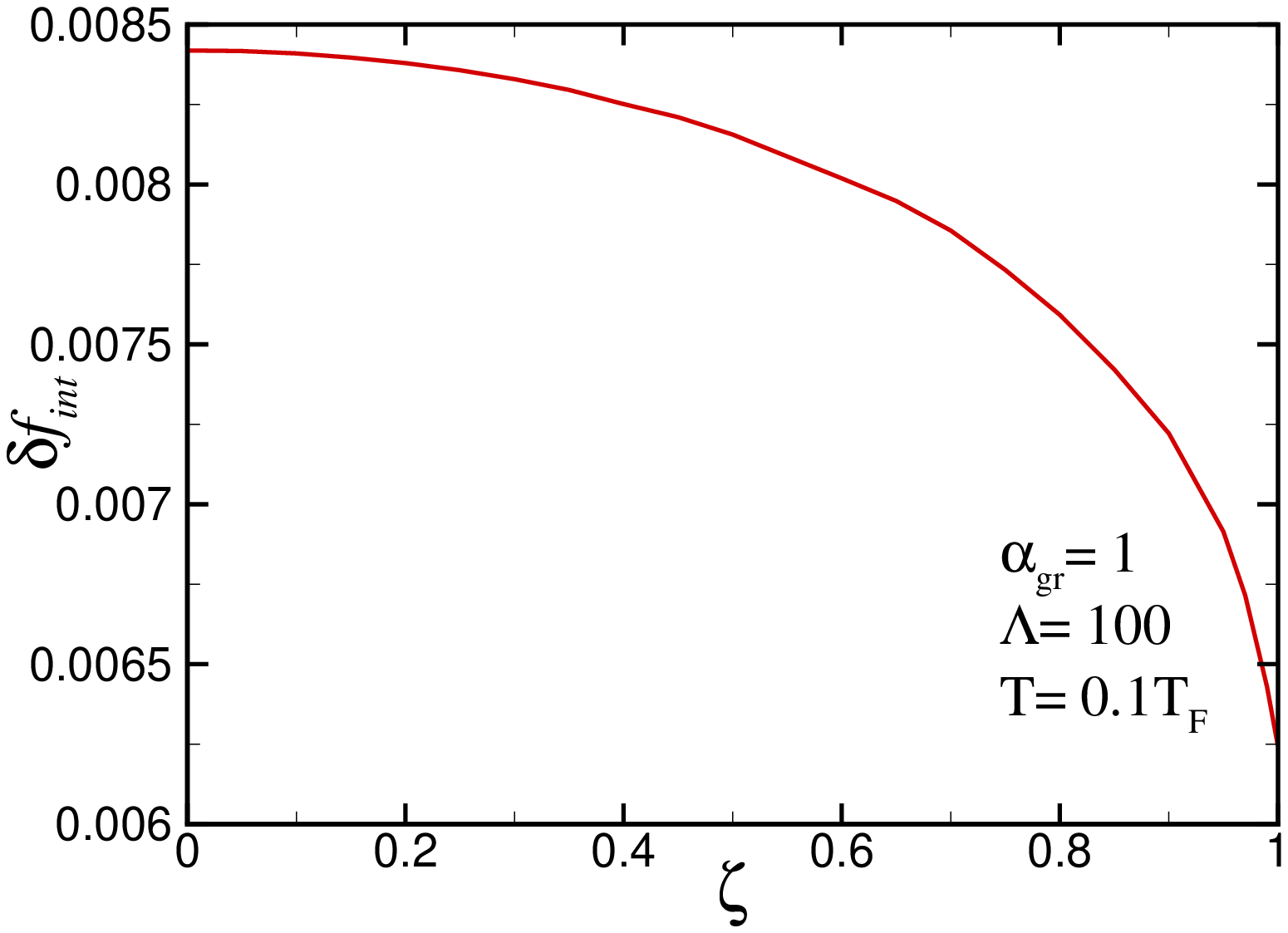}
\includegraphics[width=0.90\linewidth]{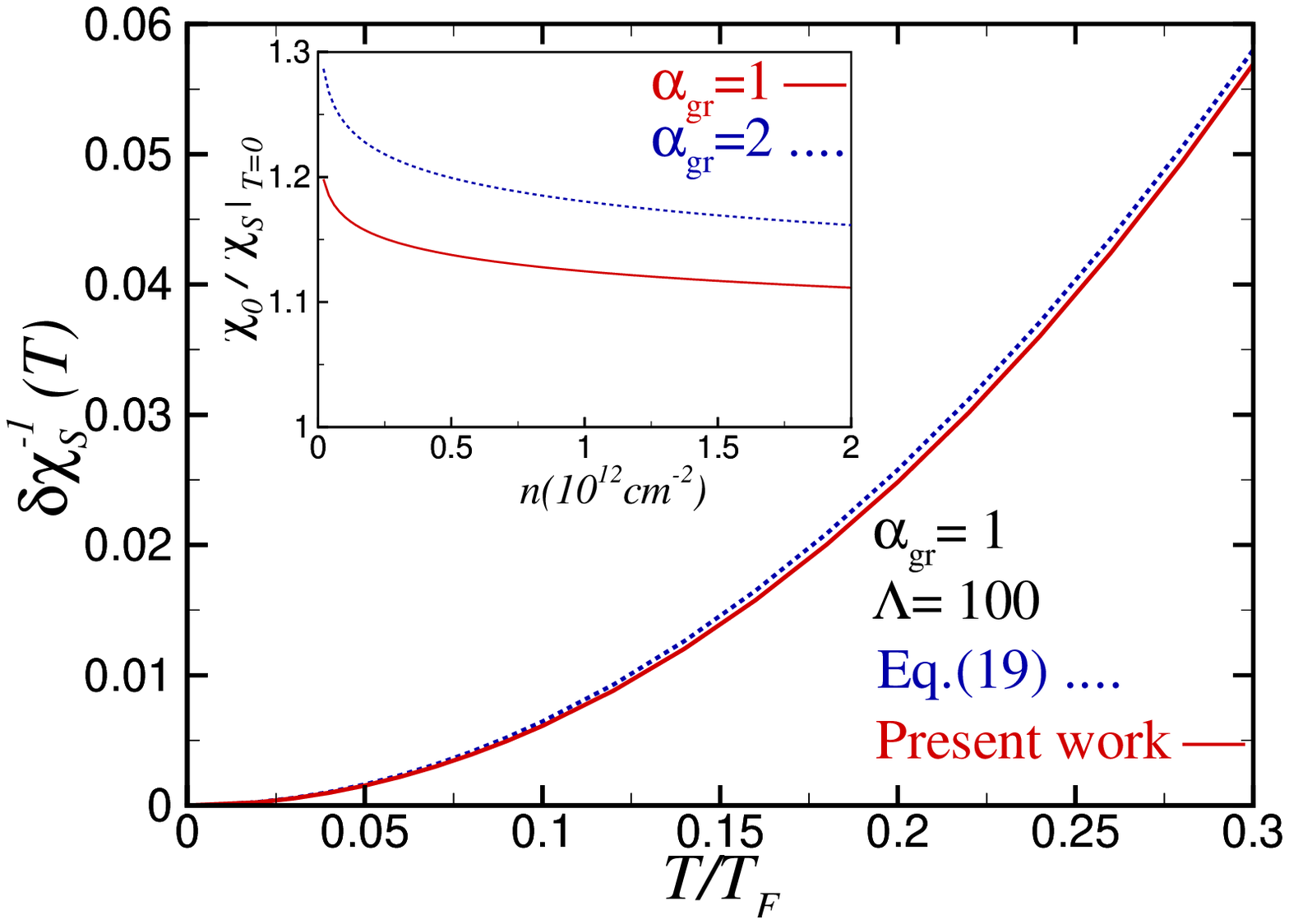}
\caption{(Color online) Upper: $\delta f_{\rm int}$ in units of $\varepsilon_{\rm F}$ as a
function of spin polarization parameter, $\zeta$ for
$\alpha_{\rm gr}=1$ and $T=0.1 T_{\rm F}$. Bottom : Numerical calculated $\delta  \chi^{-1}_s(T)=\chi^{-1}_s(T)-\chi^{-1}_s(T=0)$ in units of $\varepsilon_{\rm F}/n \mu_B^2$ as a function of temperature for $\alpha_{gr}=1$ in
comparison with the low temperature approximated expression given by Eq.(\ref{eq:chi_s}). These numerical results confirm the validity of our
analytic result for $\delta f_{\rm int}$. In the inset, the inverse spin susceptibility scaled by non-interacting one as a function of electron density in units of $10^{12}$ cm$^{-2}$ at zero temperature for different $\alpha_{gr}$ values. }
\end{center}
\end{figure}

The coupling constant integration in Eq.~(\ref{eq:int_free_energy}) can be carried out partly analytically
due to the simple RPA expression Eq.~(\ref{eq:chi_RPA}). We find that the interaction contribution to the free energy per particle $f_{\rm int}(T)$ is given by
\begin{eqnarray}\label{eq:int_free_energy_integrated}
&&f_{\rm int}(T) =\frac{1}{2}
\int\frac{d^2{\bm q}}{(2 \pi)^2}\left\{-\frac{1}{\pi n}\int_0^{+\infty}d\omega\coth{(\beta\omega/2)}\times\right.\nonumber\\
&&\left.\arctan\left[\frac{v_q[\Im m ~\chi^{\uparrow}_0+\Im m ~\chi^{\downarrow}_0]}{1-v_q[\Re e~\chi^{\uparrow}_0+\Re e~\chi^{\downarrow}_0)]}\right]-v_q\right\}\nonumber\\
&+&\frac{1}{2n}\int\frac{d^2{\bm q}}{(2 \pi)^2}\int_0^1
\frac{d\lambda}{\lambda}\coth{(\beta\omega_{\rm pl}/2)}
[\Re e~\chi^{\uparrow}_0+\Re e~\chi^{\downarrow}_0]\nonumber\\
&\times&\left|{\frac{\partial[\Re e~\chi^{\uparrow}_0+\Re
e~\chi^{\downarrow}_0]}{\partial \omega}}\right|^{-1}_{\omega =
\omega_{\rm pl}}\,.
\end{eqnarray}
In this equation the first term comes from the smooth
electron-hole contribution to $\Im m~\chi^{(\lambda)}_{\rho\rho}$,
while the second term comes from the plasmon contribution;
$\omega^{\sigma}_{\rm pl}=\omega_{\rm pl}(q,T,\lambda)$ is the plasmon
dispersion relation at coupling constant $\lambda$ which can be
found numerically by solving the equation $1-\lambda v_q \Re
e~\chi^{\sigma}_{0}(q,\omega,T)=0$. Note that in a standard 2D EG the
exchange energy starts to matter little for $T$ of order $T_{\rm
F}$ because all the occupation numbers are small and the Pauli
exclusion principle matters little. In the graphene case however
exchange interactions with the negative energy sea remain
important as long as $T$ is small compared to $v_{\rm F} k_{\rm
c}/k_{\rm B} = T_{\rm F}\Lambda$.

The free energy calculated according to Eq.~(\ref{eq:int_free_energy_integrated})
is divergent since it includes the interaction energy of the model's infinite sea of negative energy particles.
Following Vafek~\cite{vafek_prl_2007},
we choose the free energy at $T=0$, $f(T=0)$,
as our ``reference" free energy, and thus introduce the regularized quantity
$\delta f\equiv f(T)-f(T=0)$. This again can be decomposed into the sum of a noninteracting contribution,
$\delta f_0(T \to 0)=-g_v\varepsilon_{\rm F}\pi^2 (T/T_{\rm F})^2 Z(\zeta, 1/2)/12$, where $Z(\zeta,m)=(1+\zeta)^m+(1-\zeta)^m$ and an interaction-induced
contribution $\delta f_{\rm int}(T)=f_{\rm int}(T)-f_{\rm int}(T=0)$, which can be calculated from Eq.~(\ref{eq:int_free_energy_integrated}). Note that we have $f_0(T=0)=g_v\varepsilon_{\rm F} Z(\zeta,3/2)/6$.

The low-temperature behavior of the interaction contribution to the free energy can be extracted analytically with some patience.
After some lengthy but straightforward algebra we find, to leading order in $\Lambda$,

\begin{eqnarray}\label{eq:crucial}
&&\delta f_{\rm int}(T\to 0)= \varepsilon_{\rm F} \frac{\pi^2}{6}
\left(\frac{T}{T_{\rm F}}\right)^2\frac{\alpha_{\rm gr}[1-\alpha_{\rm gr}\xi(\alpha_{\rm
gr})]}{8g_{\rm v}}\nonumber\\
&&\times Z(\zeta,1/2) \ln{\Lambda}+{\rm R.~T.}
\end{eqnarray}
where the function $\xi(x)$, defined as in Eq.~(14) of Ref.~[\onlinecite{polini_ssc_2007}], is given by $\xi(x)=128/(\pi^2 x^3)- 32/(\pi^2 x^2)+ 1/x-h(\pi x/8)$,
with

\begin{equation}
h(x)=\left\{
\begin{array}{ll}
{\displaystyle \frac{1}{2x^3\sqrt{1-x^2}}\arctan{\left(\frac{\sqrt{1-x^2}}{x}\right)}} & {\displaystyle {\rm for}~x<1}\vspace{0.1 cm}\\
{\displaystyle \frac{1}{4x^3\sqrt{x^2-1}}\ln{\left(\frac{x+\sqrt{x^2-1}}{x-\sqrt{x^2-1}}\right)}} & {\displaystyle {\rm for}~x>1}
\end{array}
\right.\,.
\end{equation}
The symbol ${\rm R.~T.}$ in the l.f.s. of Eq.~(\ref{eq:crucial}) indicates regular terms, {\it i.e.} terms that, by definition, are finite in the limit $\Lambda \to \infty$.  Eq.~(\ref{eq:crucial}) represents the second important result of this work.

\begin{figure}\label{fig:three}
\begin{center}
\includegraphics[width=0.9\linewidth]{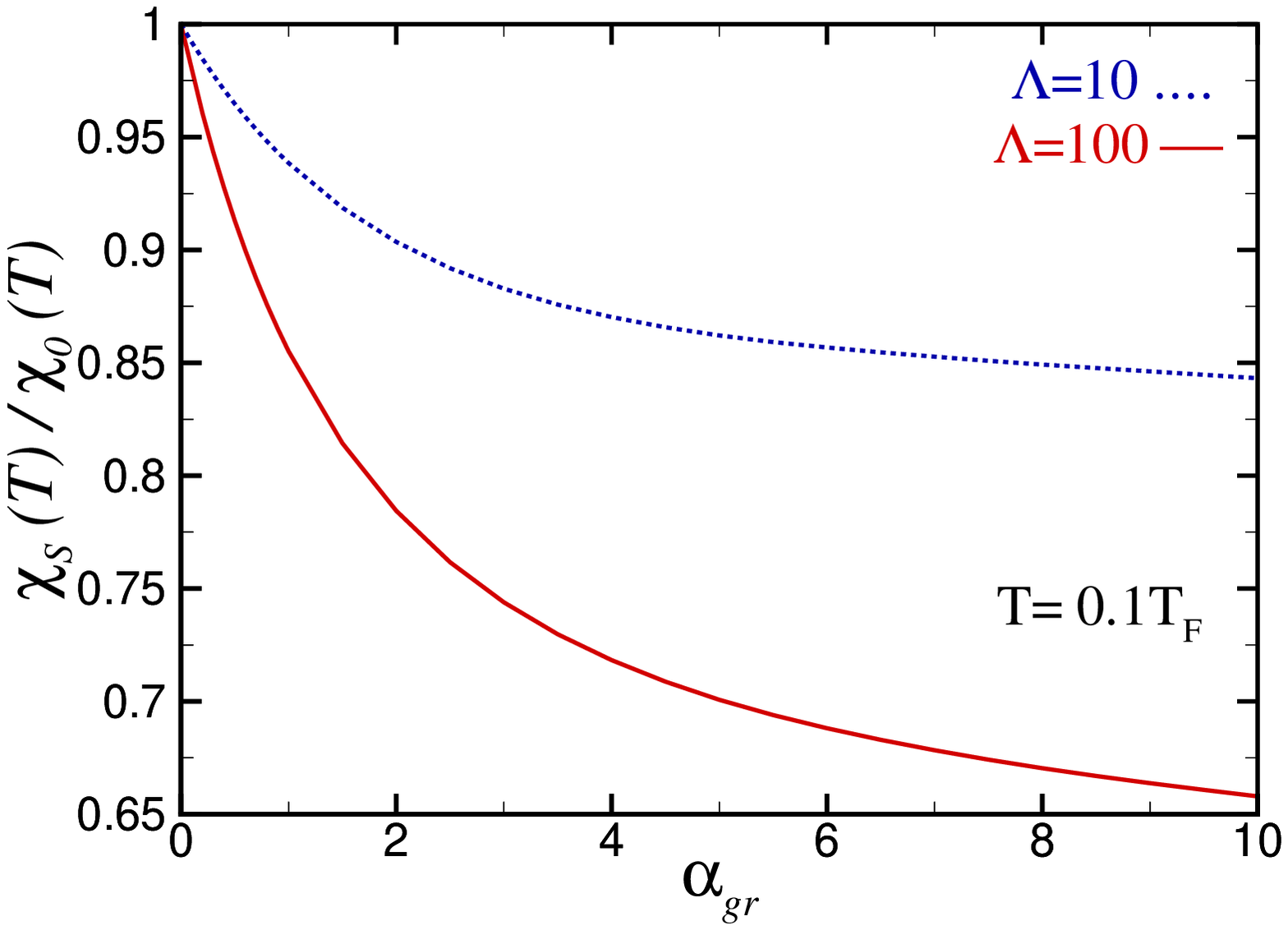}
\includegraphics[width=0.9\linewidth]{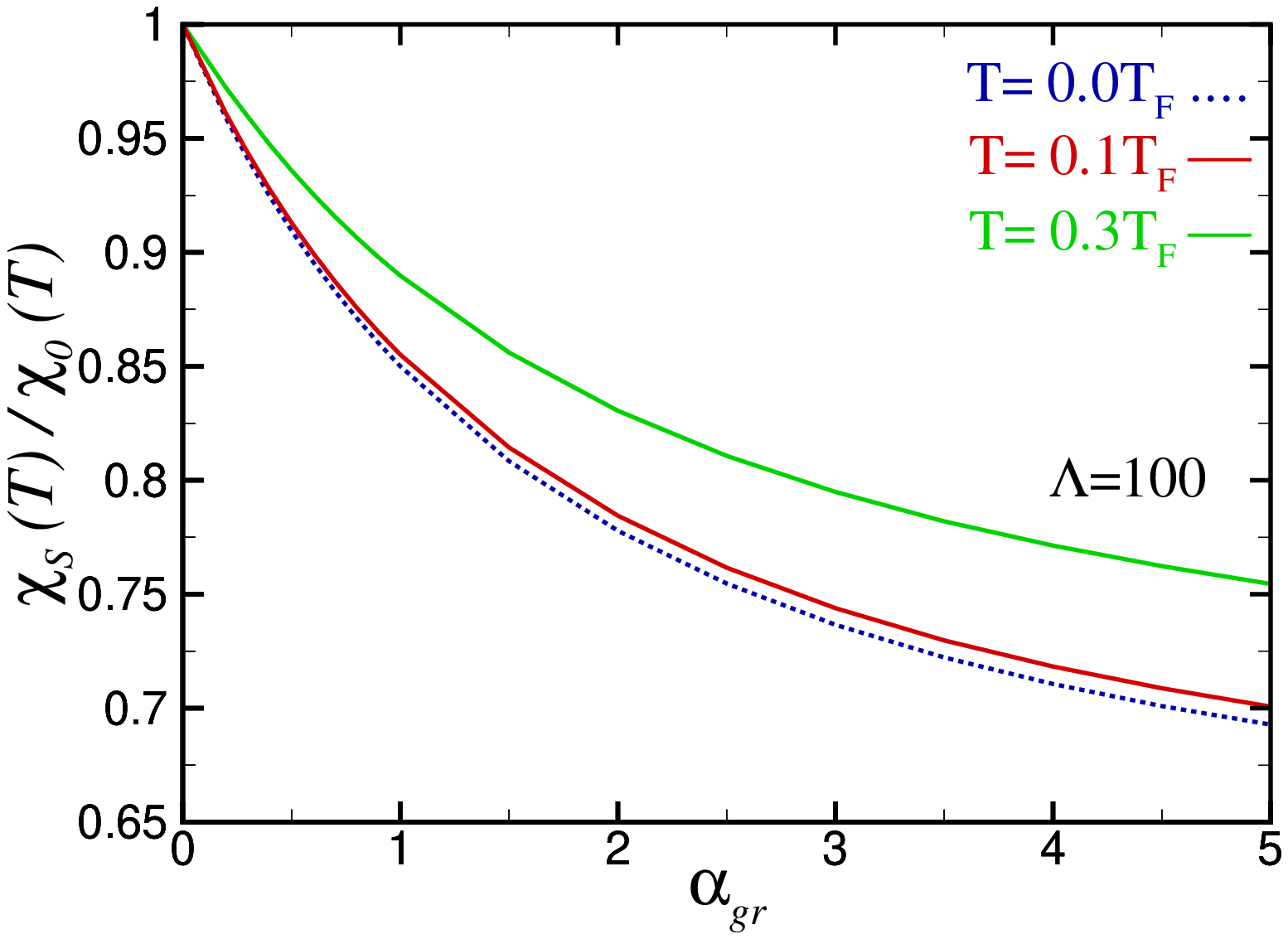}
\caption{(Color online) Upper: the spin susceptibility (in units of the non-interacting spin susceptibility $\chi_0$)
as a function of coupling constant for two values of ultraviolet cut-off, $\Lambda=10$ and $100$.
The spin susceptibility decreases with increasing $\Lambda$ or the coupling constant.
Bottom: the same as upper panel for three values of temperature for $\Lambda=100$.}
\end{center}
\end{figure}
We thus see that $\delta f_{\rm int}(T\to 0) \propto T^2$ in
Eq.~(\ref{eq:crucial}) implies a conventional Fermi-liquid
behavior with a linear-in-$T$ specific heat. We are thus led to
conclude, in full agreement with the zero-temperature calculations
of the quasiparticle energy and lifetime performed in
Refs.~\cite{polini_ssc_2007,polini_prb_2008}, that doped graphene
sheets are normal Fermi liquids.

It would be worthwhile obtaining the high-temperature dependence of $\delta f$. Since we are always measuring energies in units of the Fermi energy, therefore our high energy results are relevant to Dirac point physics. The undoped limit for us is the limit of vanishing the Fermi energy. To obtain the results for undoped graphene, let's consider the paramagnetic case. By replacing the Fermi energy with $k_{\rm B} T$ and therefore, $k_{\rm F}$ with $k_{\rm B}T/\hbar v_{\rm F}$ in Eq.~\ref{eq:crucial}, the correction to the free-energy is given by $\delta f_{\rm int}(T\gg T_{\rm F}, n \simeq 0) \sim T^3 \alpha_{gr}[1-\alpha_{gr}\xi(\alpha_{gr})] \ln(k_{c}/T) /(8 g_{v})$. Importantly, this expression, apart from a constant~\cite{note}, is coincide with that result obtained in Eq. 13 by Vafek~\cite{vafek_prl_2007}. Therefore, we expect that in the limit that $T\gg T_{\rm F}$ and for every $\zeta$ value, the temperature dependence of the free-energy correction behaves like $T^3 \ln(k_{c}/T)$.

The spin susceptibility, on the other hand, can be calculated from
the second derivative of the Helmholtz free energy and it reads
\begin{equation}
\frac{1}{\chi_s(T)}=\frac{1}{n\mu^2_B} \frac{\partial^2
[f_0(T,\zeta)+ f_{\rm int}(T,\zeta)]}{\partial \zeta^2}~|_{\zeta=0}~,
\end{equation}

where $\mu_B$ is the Bohr magneton~\cite{footnote}.

It is easy to calculate the non-interaction spin susceptibility and it turns out that
\begin{equation}\label{eq:chi}
\chi_0^{-1}(T)=\frac{1}{n\mu^2_B}\frac{g_v \varepsilon_{\rm F} }{4 }\{1+\frac{\pi^2}{6}(\frac{T}{T_{\rm F}})^2\}~.
\end{equation}

At low temperature, by using $\delta f_{\rm int}$ to leading order in $\Lambda$, the temperature dependence of the correction to the spin susceptibility is thus given by

\begin{eqnarray}\label{eq:chi_s}
\delta \chi_s^{-1}(T)&=&\chi^{-1}_s(T)-\chi^{-1}_s(T=0)\nonumber\\
&=&\frac{\varepsilon_{\rm F} \pi^2}{8 n\mu_B^2}(\frac{T}{T_{\rm F}})^2\left[ \frac{g_v}{3}-\eta \ln \Lambda \right]
\end{eqnarray}

where $\eta=\alpha_{gr}(1-\alpha_{gr} \xi(\alpha_{gr}))/12 g_v$. It is obvious from the expression
that $\chi_s(T) \propto T^{-2}$ at low temperature limit. This expression represents
our important result in this work.

\section{NUMERICAL RESULTS}

In this section, we present the most important results of the spin susceptibility in doped graphene sheets by using mentioned formalism.

The semi-analytical expressions for $\Re
e~\chi^{\sigma}_{0}(q,\omega,T)$ and $\Im
m~\chi^{\sigma}_{0}(q,\omega,T)$ constitute the first important
result of this work. In Fig.~1 we have plotted the
major part of the dynamic response function as a function of $q/k_{\rm F}$ for
different values of $\zeta$. Sharp cutoffs in the imaginary part of
$\chi^{\uparrow}_{0}(q,\omega,T)$ are related to
the rapid swing in the real part of $\chi^{\uparrow}_{0}(q,\omega,T)$. These
behaviors are in result of the fact that the real and imaginary
parts of the polarization function are related through the Kramers-Kr\"{o}nig
relations. Importantly, the sign change of the real part from negative to
positive shows a sweep across the electron-hole continuum. It is important to note that there is a non-monotonic behavior of $\chi^{\sigma}_{0}(q,0,T)$ as a temperature dependent originates from a competition between intra- and
inter-band contributions to this quantity~\cite{ramezan}. However, the
spin polarization parameter dependence of the Lindhard function is a monotonic behavior at any frequency.

Fig.~2 (upper) shows the interaction contributions of the free energy as a
function of spin polarization for $T=0.1 T_{\rm F}$. Our numerical results
show that the contribution of electron-electron interactions in the free energy
decrease by increasing the spin polarization. It changes slightly at low $\zeta$
values and sharply decreases near the ferromagnetic case. The reason is that the exchange and correlation energies mostly change around the fully ferromagnetic point. Moreover, the slope of exchange and correlation energies with respect to $\zeta$ around $\zeta=1$ have opposite signs~\cite{QA} and the interaction contribution tends to the correlation sign at certain value of $\zeta$. Fig.~2 (bottom) shows
the numerical calculated $\chi^{-1}_s(T)-\chi^{-1}_s(T=0)$ as a function of temperature in
comparison with that result obtained at low temperature and leading order of $\Lambda$.
We can easily see that those results are very close at low temperature and confirm
that the spin susceptibility behaves as $T^{-2}$ in this region. In addition, this comparison allows us to use the approximated analytical expression given by Eq.~(\ref{eq:chi_s}) for the temperature correction of the spin susceptibility till $T \leq 0.3 T_{\rm F}$. Furthermore, we can see that the spin susceptibility sharply decreases with increasing temperature. On the other hand, temperature dependence of the spin susceptibility is in contrast to that result obtained for the diamagnetic undoped graphene sheet. To seek comprehensive study, we have numerically calculated $\chi_0/\chi_s$ as a function of electron density in units of $10^{12}$ cm$^{-2}$ at zero temperature, $T=0$. Our results are shown in the inset of Fig.~2 (bottom) and show that the spin susceptibility increases by increasing the electron density~\cite{barlas_prl_2007}.

Finally, we show the spin susceptibility scaled by its
non-interacting value as a function of the coupling constant for
(upper panel) two values of the ultraviolet cut-off and (bottom panel) different
values of temperatures in Fig.~3. These results are obtained numerically by taking the full terms of Eq.~(\ref{eq:int_free_energy_integrated}). We can clearly see that the spin
susceptibility increases by increasing the electron density ( or
decreasing the $\Lambda$ values) while it decreases by increasing
the interactions at certain temperature value. The reason is that the exchange contribution term makes a positive contribution to Eq.~(\ref{eq:chi}), thus tending to reduce the spin susceptibility (with respect to its noninteracting value), again in contrast to what happens in the standard 2D EG where exchange enhances the spin susceptibility. The correlation term instead makes a negative contribution to Eq.~(\ref{eq:chi}), thus tending to enhance the spin susceptibility. In the 2D EG, correlations tend to reduce the spin susceptibility.

\section{Conclusions}

We have presented semi-analytical expressions for the real and the
imaginary parts of the resolved spin dependence of density-density
linear-response function of noninteracting massless Dirac fermions
at finite temperature. These results are very useful in order to study
finite-temperature screening within the Random Phase
Approximation. For example they can be used to calculate the spin
dependence of the conductivity at finite temperature within
Boltzmann transport theory .

The Lindhard function at finite temperature is also extremely
useful to calculate finite-temperature equilibrium properties of
interacting massless Dirac fermions, such as the specific heat and
the compressibility. For example, in this work we have been able
to show that, at low temperatures, the paramagnetic spin
susceptibility of interacting massless Dirac fermions behaves like
$T^{-2}$ at low temperature. Even though the charge and spin
susceptibilities behave similarly at zero temperature, their temperature
dependencies are totally different. We have obtained an analytical expression for the spin susceptibility in the leading order of cut-off and showed that one can use that in the low temperature range for experimental access.

We remark that in a very small density region, the system is highly
correlated and a model going beyond the RPA is necessary to account for increasing correlation effects
at low density.

\section{Acknowledgement}
We thank A. Naji for his useful comments. We would like to dedicate this report to the memory of my sister, " Farideh Asgari".


\begin{thebibliography}{77}
\bibitem{footnote}
  The second
derivative is calculated using the full temperature dependent
free-energy of the non-interacting system, $f_0(T)$, and
not its analytical expression reported above that is valid only
for $T \ll T_{\rm F}$.
\end{thebibliography}
\end{document}